\documentclass[iop]{emulateapj}

\newcommand\curf{{\cal F}}
\shorttitle{Stellar parameters and accretion rate of the transition disk star HD 142527 from X-Shooter.}
\shortauthors{Mendigut\'\i{}a et al.}

\begin{document}


\title{Stellar parameters and accretion rate of the transition disk star HD 142527 from X-Shooter.}

%\author{Authors}
%\affil{Affiliations}
%\email{email}

\author{I. Mendigut\'\i{}a} 
\affil{School of Physics \& Astronomy, University of Leeds, Woodhouse
  Lane, Leeds LS2 9JT, UK}
\email{I.Mendigutia@leeds.ac.uk}

\author{J. Fairlamb}
\affil{School of Physics \& Astronomy, University of Leeds, Woodhouse
  Lane, Leeds LS2 9JT, UK}

\author{B. Montesinos}
\affil{Centro de Astrobiolog\'{\i}a, Departamento de Astrof\'{\i}sica
  (CSIC-INTA), ESAC Campus, P.O. Box 78, 28691 Villanueva de la
  Ca\~nada, Madrid, Spain.}

\author{R.D. Oudmaijer} 
\affil{School of Physics \& Astronomy,
  University of Leeds, Woodhouse Lane, Leeds LS2 9JT, UK}

\author{J.R. Najita} 
\affil{National Optical Astronomy Observatory,
  950 N. Cherry Avenue, Tucson, AZ 85719, USA}

\author{S.D. Brittain} 
\affil{Department of Physics and Astronomy,
  Clemson University, Clemson, SC 29634-0978, USA}

\author{M.E. van den Ancker} 
\affil{European Southern Observatory,
  Karl-Schwarzschild-Str. 2, D-85748 Garching b. M\"unchen, Germany}


%\altaffiltext{1}{Notita}
%\altaffiltext{2}{Society of Fellows, Harvard University.}
%\altaffiltext{3}{present address: Center for Astrophysics,
%    60 Garden Street, Cambridge, MA 02138}
%\altaffiltext{4}{Visiting Programmer, Space Telescope Science Institute}
%\altaffiltext{5}{Patron, Alonso's Bar and Grill}


% >
% > BM modification follows
% >

\begin{abstract}

HD 142527 is a young pre-main sequence star with properties indicative
of the presence of a giant planet or/and a low-mass stellar
companion. We have analyzed an
X-Shooter/Very Large Telescope spectrum to provide accurate stellar
parameters and accretion rate. The analysis of the spectrum, together with constraints
provided by the SED fitting, the distance to the star (140 $\pm$
  20 pc) and the
use of evolutionary tracks and isochrones, lead to the following set
of parameters $T_{\rm eff}$= 6550 $\pm$ 100 K, $\log g$ = 3.75 $\pm$ 0.10,
$L_*/L_\odot$ = 16.3 $\pm$ 4.5, $M_*/M_\odot$ = 2.0 $\pm$ 0.3 and an age of
5.0 $\pm$ 1.5 Myr. This stellar age provides further constrains to the mass of the possible companion estimated by Biller et al. (2012), being in-between 0.20 and 0.35 M$_{\odot}$. Stellar
accretion rates obtained from UV Balmer excess modelling, optical
photospheric line veiling, and from the correlations with several
emission lines spanning from the UV to the near-IR, are consistent to
each other. The mean value from all previous tracers is 2 ($\pm$
1) $\times$ 10$^{-7}$ M$_{\odot}$ yr$^{-1}$, which is within the
upper limit gas flow rate from the outer to the inner disk recently
provided by Cassasus et al. (2013). This suggests that almost all gas
transferred between both components of the disk is not trapped by the
possible planet(s) in-between but fall onto the central star, although
it is discussed how the gap flow rate could be larger than previously
suggested. In addition, we provide evidence showing that the stellar
accretion rate of HD 142527 has increased by a factor $\sim$ 7 on a
timescale of 2-5 years. 
\end{abstract}


\keywords{Accretion, accretion disks --- circumstellar matter ---
  planet--disk interactions --- protoplanetary disks --- stars:
  fundamental parameters --- stars: pre-main sequence}


\vspace*{5mm}

\section{Introduction}
\label{Section:Introduction}

Since the first discovery of a planet orbiting a star other than the
Sun \citep{MayorQueloz95}, around 1000 exoplanets have been confirmed
around main sequence stars with different properties. Indeed, the
latest results from Kepler are consistent with nearly every Sun-like
star in the galaxy harboring a planet \citep{Fressin13}. Planet
formation is expected to occur in the gas and dust disks that surround
young stars during the first $\sim$ 10 Myr. However, the detection of
planets caught in their process of formation is limited only to a few
candidates
\citep[e.g.][]{Huelamo11,Kraus12,vanEyken12,Brittain13}. Nevertheless,
several pre-main sequence (PMS) stars present specific properties that
make them the ideal candidates to look for planets in their disks.

One of these objects is the isolated, intermediate-mass PMS star
\object{HD 142527}, which shows several observational signatures of
ongoing gas giant planet formation. Using spectra taken by the SWS and
LWS instruments mounted on the Infrared Space Observatory,
\citet{Malfait99} reported two different dust components in its almost
face-on disk. This double nature is accompanied by a specific
morphology. Near and mid-infrared images show that the disk has a
large inner hole, two bright facing arcs, and a spiral arm extending
from one of them \citep{Fukawaka06,Fujiwara06}. \citet{Verhoeff11}
provided a more precise description of the circumstellar environment
of HD 142527, constituted by a geometrically thin inner disk that
extends from 0.3 to 30 AU, an optically thin halo-like component of
dust in the inner disk regions, and a massive outer disk running from
130 AU up to 200 AU. These authors find that the peculiar properties
of this system are indicative of on-going planet formation \citep[see
  also][]{Rameau12}. \citet{Casassus12} proposed that the gap between
the inner and outer disk could have been carved out by a giant planet
located at around $\sim$ 90 AU, which causes perturbations in the
whole disk and the appearance of spiral arms. More recently,
\citet{Casassus13} used ALMA observations to report the presence of
planet-induced CO gas channels flowing from the outer to the inner
disk at a rate ranging between 7 $\times$ 10$^{-9}$ and 2 $\times$
10$^{-7}$ M$_{\odot}$ yr$^{-1}$. This flow would supply material to
the inner disk and is consistent with a previous estimate of the
stellar accretion rate \citep[][see below]{GarciaLopez06}. Other
possible scenarios explaining the peculiar properties of the HD 142527
include the possible presence of a stellar companion
\citep{Fukawaka06,Baines06,Biller12,Close14}.

While observations of HD~142527 have generated a great deal of
excitement, the stellar and accretion properties are poorly
characterized. Based on unpublished spectra and photometry, HD 142527
was classified as a Herbig Ae/Be (HAeBe) star by
\citet{Waelkens96}, and a spectral type F6-F7III is
the most commonly assumed \citep{Houk78,vandenancker98}. From an
Hipparcos distance of $\sim$ 200 pc, \citet{vandenancker98} derived a
stellar mass and luminosity of 3.5 M$_{\odot}$ and 69 L$_{\odot}$, a
surface temperature of 6300 K and an age of 10$^5$ years. These values
are partially similar to the ones obtained in subsequent works
\citep{vanboekel05,Manoj06}. In contrast, \citet{Fukawaka06} assumed a
distance of 140 pc, stellar mass of 1.9 M$_{\odot}$, and an age of 2
Myr. Recent works \citep[e.g.][]{Biller12,Rameau12} assume the
parameters provided by \citet{Verhoeff11}: d = 145 pc, M = 2.2
M$_{\odot}$, L = 20 L$_{\odot}$, T = 6250 K, age = 5 Myr, R = 3.8
R$_{\odot}$. Regarding the stellar accretion rate, there are only two
estimates in the literature. \citet{GarciaLopez06} provided $\sim$ 7
$\times$ 10$^{-8}$ M$_{\odot}$ yr$^{-1}$ from the correlation with the
Br$\gamma$ luminosity in \citet{Calvet04}. More recently,
\citet{Salyk13} obtained a correlation between the accretion
luminosity derived from different observational methods, and the
\ion{H}{1} Pfund $\beta$ line luminosity. Using this correlation, a
mass accretion rate of $\sim$ 1 $\times$ 10$^{-7}$ M$_{\odot}$
yr$^{-1}$ was derived for HD 142527. However, accretion rates obtained
from individual lines suffer from large uncertainties, while the
average derived from several line luminosities measured simultaneousy
is a more robust estimate \citep{Rigliaco12}. Moreover, accretion
variability has been reported for several HAeBes
\citep{Mendi11,Mendi13}, for which estimates obtained at different
epochs become necessary.

The parameters characterizing the central star determine the energy
received by the disk and influence its geometry, energy balance,
accretion rate and contribution to the spectral energy
distribution. It is our aim to provide an independent, self-consistent
study of the stellar parameters of HD 142527, that will serve for
future detailed analysis of its disk properties. We make use of an
X-Shooter/{\it Very Large Telescope} \citep[VLT;][]{Vernet11}
spectrum, whose wide wavelength coverage from the UV to the near-IR
provides an accurate estimate of the mass accretion rate from several
tracers observed simultaneously.

This paper is organized as follows. Section \ref{Section:data}
describes the X-Shooter spectrum used in this
work. The stellar properties of HD 142527 are derived in
Sect. \ref{section:stellarparam}, and the accretion rate in
Sect. \ref{section:accretion}. Results are summarized and discussed in
Sect. \ref{section: conclussions}.

\section{Observational data}
\label{Section:data}

This work is mainly based on the spectrum of HD 142527 taken on 2010-04-01 UT08:43, using the X-Shooter spectrograph mounted on the VLT
\citep{Vernet11}. X-Shooter covers a wide spectral range (300nm-2.5
$\mu {\rm m}$) across three arms: 300-590 nm, 530 nm-1.0 $\mu {\rm
  m}$, and 1.0-2.4 $\mu {\rm m}$ with the UVB, VIS and NIR arms,
respectively. The smallest slit widths available were used: 0.5''
(UVB), 0.4'' (VIS), and 0.4'' (NIR). This gives the highest spectral
resolution achieveable with X-Shooter; R=9100, 17400 and 11300,
respectively. The target was observed in nodding mode with a total
exposure time of 82 seconds.

The spectrum was reduced in two stages. The first follows the standard
reduction from the X-Shooter pipeline v0.9.7
\citep{modigliani2010}. The pipeline-reduced 1D spectrum is suitable
for measurement of spectral lines but not for an analysis of the
continuum shape, due to problems encountered when the separate echelle
orders are merged. Therefore, a second stage of reduction was required
to derive the proper SED shape. The flux standard observed on the same
night has a poor S/N ratio and is rejected in favour of a higher S/N
telluric standard, observed shortly after the target star. The exact
flux of the telluric standard is not known and therefore cannot
produce an accurate flux correction. However, absolute flux
calibration is not required for this work, which only needs from an
accurate determination of the SED shape in the UVB arm
(Sect. \ref{section:balmer}). In order to perform this correction,
each echelle order of the telluric standard was corrected individually
by comparing against a model of the same spectral type. We obtained
accurate stellar parameters for our telluric standard, HIP 77968, by
comparing the wings of the Balmer re-combination lines with a grid of
Kurucz-Castelli model atmospheres from \citet{munari2005}. The Balmer
wings have been shown to be sensitive to change in both ${\rm T}_{\rm
  eff}$ and log g \citep{montesinos2009}. Using this method, we
derived ${\rm T}_{\rm eff} = 12000 \pm 250$ K and log g = 4.2 $\pm$
0.15 for the telluric star. These values agree within the spectral
type range of the literature \citep[e.g.][]{straizys1981}, and were
used for the corrections. Each echelle order of the telluric spectrum
was divided through by the corresponding Kurucz model atmosphere to
produce a correction function. This was then applyed to the 1D echelle
orders of HD 142527, resulting in each echelle order now having the
correct SED shape. They were then merged together where they overlap
to produce the corrected 1D spectra.

\section{Stellar parameters}
\label{section:stellarparam}

The problem of the distance to this star is fundamental to determine the stellar parameters, which in turn effects the accretion rate estimates (Sect. \ref{section:accretion}). The parallax from the Hipparcos catalogue is $\pi\!=\!5.04\pm 1.18$ mas, which implies a distance of $198.4_{-37.6}^{+60.7}$ pc; \citet{vanLeeuwen07} gives a smaller value for the parallax, namely $\pi\!=\!4.29 \pm 0.98$ mas (d = $233.1_{-43.4}^{+69.9}$ pc). However, in both cases $\sigma_\pi/\pi>0.17$, therefore the Lutz-Kelker bias \citep{Lutz73} prevents from deducing a reliable distance to the star from the parallaxes. The consensus is to adopt a shorter distance of 140 ($\pm$ 20) pc \citep{Fukawaka06}, which is the one adopted in this work. There are strong reasons supporting that assumption: the star is within $\sim\!1$ degree of the Lupus IV cloud (at a mean distance of 140 pc) and has known Lupus and Sco-Cen members in its vicinity \citep[the][catalog shows four Sco-Cen members within 2 degrees, and nine within 3 degrees]{deZeeuw99}. In addition, the proper motion and radial velocity of HD 142527 are similar to those of other of stars in the Lupus clouds \citep[see][]{vanLeeuwen07,Galli13}.

A visual inspection of the blue part of the X-Shooter spectrum and a comparison
with MK standars \citep[see e.g.][]{Gray09} hints that the spectral
type is between $\sim$F3 and F8, whereas the luminosity class is
between V and III. However, the photospheric lines that are
primary indicators for the stellar temperature in F stars, namely the Balmer hydrogen
lines, cannot be used here since their profiles show a non-photospheric component caused by accretion (Sect. \ref{Section: lines}). For the same reason, the \ion{Ca}{2} HK and \ion{Na}{1}D lines, as well as \ion{Ca}{1} at 4226 \AA{} and \ion{Fe}{1} at 4046, 4383 \AA{} 
\citep[which are also
temperature indicators in combination with H$\gamma$ and H$\delta$;][]{Gray09}, cannot be used either. In addition, the spectral
resolution --lower than that of a echelle spectrum-- prevents from applying diagnostics based
purely on line ratios. Therefore an alternative was followed to further constrain the stellar parameters.

\begin{figure}
%[!hbtp]
\centering
 \includegraphics[width=86mm,clip=true]{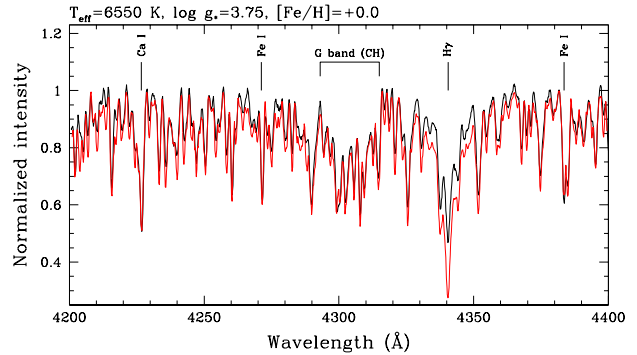}
\caption{Observed spectrum of HD 142527 (black) in the interval
  4200-4400 \AA{} and best fit synthetic spectrum (red, see text) computed with T$_{\rm
    eff}$=6550 K, $\log g_*$=3.75, solar metallicity and broadened
  with $v \sin i$=45 km/s \citep{Glebocki05}.}
\label{Figure:synthesis}
\end{figure}

\begin{figure}
%[!hbtp]
\centering
 \includegraphics[width=86mm,clip=true]{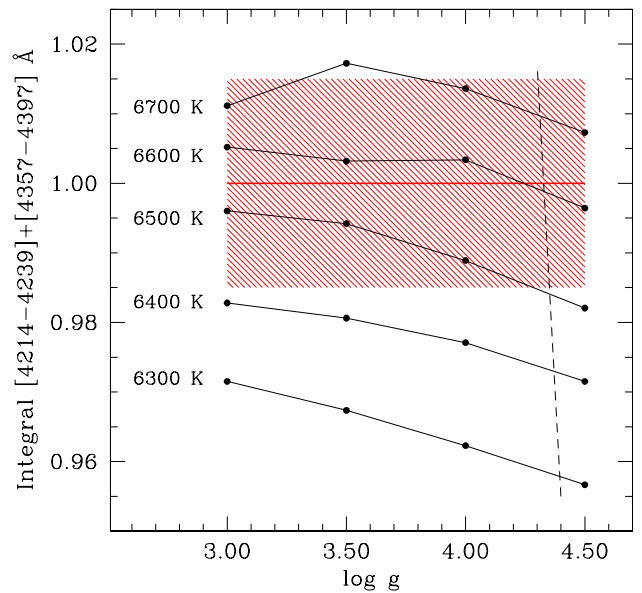}
\caption{Integrals of the area under the normalized intensity of the
  synthetic models (black dots and lines) in the regions [4214-4239] +
  [4357-4397] \AA{} divided by the area under the intensity of the
  observed spectrum in the same region (thick red line), that has been
  scaled to 1.0. The shaded region marks the uncertainty in the latter
  integral due to potencial inaccuracies in the normalization of the
  observed spectrum. The dashed black line marks the values of $\log
  g$ for main-sequence stars. See text for details}
\label{Figure:integrals}
\end{figure}

We have used the region 4200--4400 \AA{} to obtain an estimate of the
effective temperature (Fig. \ref{Figure:synthesis}). A grid of high-resolution spectra with $T_{\rm
eff}$ = 6300--6700 K, in steps of 100 K, and $\log g$=3.0, 3.5, 4.0,
4.5 was synthetized. The programmes {\sc synthe} and {\sc atlas}
(Kurucz, 1993) fed with the models describing the stratification of
the stellar atmospheres \citep{Castelli03} and solar abundances were used. For a given gravity, the depth/equivalent width of the vast majority
of the lines decrease as the temperature increases --as expected-- so
we have used this fact to constrain the value of $T_{\rm eff}$. To
quantify that, we normalized to unity the
observed spectrum and the synthetic models, and integrated the area
below the spectral lines (avoiding the CH G-band around 4300 \AA,
which shows clear gravity effects for a given temperature, and
H$\gamma$).  This analysis has the advantage that the integrals do not
depend on the projected rotational velocity $ v \sin i$, because rotational broadening
conserves equivalent widths. The results can be found in
Fig. \ref{Figure:integrals}. The thick horizontal red line marks the
value of the area contained under the observed spectrum, that has been
scaled to 1.0, and the shaded area accounts for potential inaccuracies
in the normalization of the observed spectrum and hence in the value
of the integral; the black dots and lines joining them are the areas
contained under the corresponding synthetic models normalized to the
area under the observed spectrum. The dashed line marks the values of
$\log g$ for main-sequence stars in this range of temperatures. As it
can be seen, the effective temperature should be higher than 6400 K but not than 6700 K.

\begin{figure}
%[!hbtp]
\centering
 \includegraphics[width=86mm,clip=true]{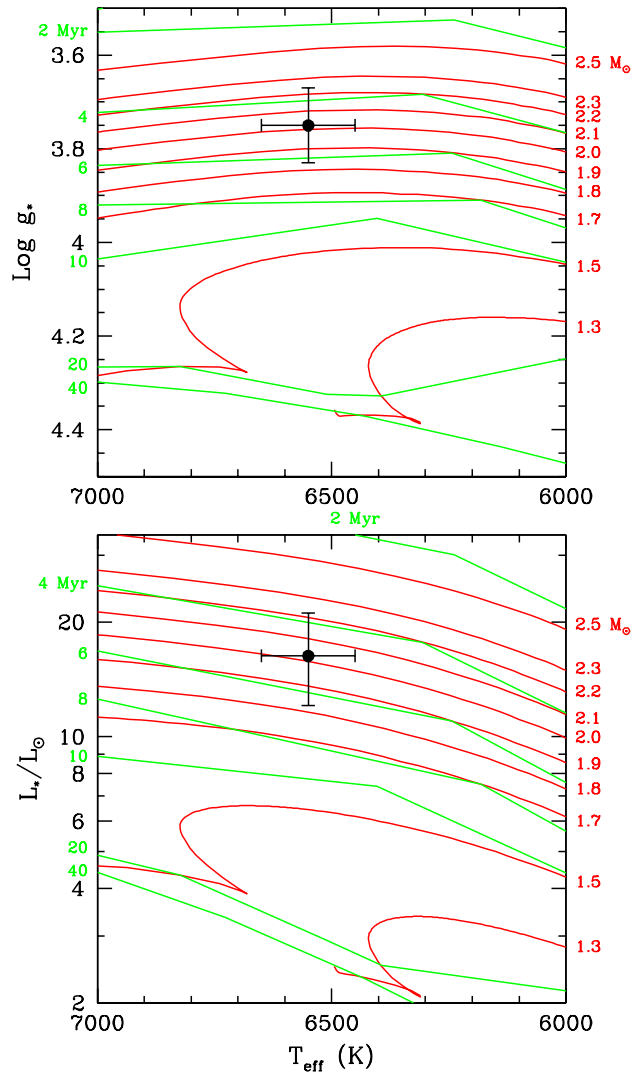}
\caption{$\log g_*$ - $T_{\rm eff}$ (top) and $L_*/L_\odot$ - $T_{\rm
    eff}$ (bottom) HR diagrams showing the position of HD 142527
  according to the set of parameters determined in this work. The
  Yonsei-Yale set \citet{Yi01} of tracks (red) and isochrones (green)
  for solar metallicity have been used.}
\label{Figure:hrdiagrams}
\end{figure}

In order to constrain the value of $\log g$, all the luminosity
criteria given by \citet{Gray09} for F-type stars have been checked
for the latter range of temperatures. Values between $\log g$=3.5 and 4.0 are
valid but unfortunately none of the criteria provide clues to break
the degeneracy and give an accurate value for the gravity. This problem was circunvented using the distance to
the star. We have taken as a baseline the model with $T_{\rm
eff}$=6550 K and $\log g$=3.75 (right in between the limits in
temperature and gravity given by the spectral analysis above). A
Kurucz low-resolution model with these parameters was
fitted to the available optical photometry (Johnson $BV$ and Cousins $I_{\rm c}$ photometry from the Hipparcos CDS
catalogue {\tt I/239/hip\_main}\footnote{from the comparison with a different set of optical measurements \citep{Malfait98}, there is no evidence for strong photometric variability in HD 142527.}).  By integrating the dereddened
fit, the observed photospheric flux can be
estimated. On the other hand, using evolutionary tracks and
isochrones (Fig. \ref{Figure:hrdiagrams}), the point (6550 K, 3.75) can be taken from the ($T_{\rm
eff}$, $\log g$) HR diagram to the ($T_{\rm eff}$, $L_*/L_\odot$) HR
diagram, allowing us to estimate the stellar luminosity. From the photospheric flux and the luminosity, the distance can be
estimated; for that particular model it turns out that $d\!=\!144$ pc,
in excellent agreement with the distance of 140 pc assumed for the
star. An uncertainty of $\pm 20$ pc, translated into the luminosities,
implies an uncertainty of $\pm 0.10$ in $\log g$.  Distances derived
for values of $\log g$ outside the range $3.75\pm 0.10$ are
inconsistent with the star being at $140\pm 20$ pc. 

The uncertainty in the temperature, set in $\pm 150$ K by the results
displayed in Fig. \ref{Figure:integrals}, can be further reduced to
$\pm 100$ K when models with temperatures outside the range $6550\pm
100$ K are compared with the observed spectrum in the region of the CH
G-band (Fig. \ref{Figure:synthesis}), which for a given $T_{\rm eff}$ shows a strong dependence with
$\log g$. For $T_{\rm eff}\!<\!6450$ and $T_{\rm
eff}\!>\!6650$ K the gravities needed to reproduce that feature are
incompatible with the star being at $140\pm 20$ pc.

Table \ref{Table:stelpar} summarizes the stellar parameters derived for HD
142527 from a combined analysis of the spectrum, SED fitting, and the use of evolutionary tracks and isochrones. These allowed us to also estimate the mass and age of the star. The colour excess in the table
corresponds to the reddening required to fit the Kurucz low-resolution
spectrum with 6550 K, $\log g$=3.75 to the observed photometry.

\begin{table}
\centering
\renewcommand\tabcolsep{2.5pt}
\caption{Stellar parameters of HD 142527 derived in this work}
\label{Table:stelpar}
\begin{tabular}{lll}
\hline\hline
\hline
Effective temperature &$T_{\rm eff}$ (K)      &6550 $\pm$ 100   \\
Stellar mass          &$M_*$ (M$_{\odot}$)    &2.0  $\pm$ 0.3   \\
Surface gravity       &$\log g_*$ (cm/s$^2$) &3.75 $\pm$ 0.10  \\
Stellar luminosity    &$L_*$ (L$_{\odot}$)    &16.3 $\pm$ 4.5   \\
Age                   &$t$ (Myr)             &5.0 $\pm$ 1.5    \\
Colour excess         &$E(B\!-\!V)$ (mag)    &0.25 $\pm$ 0.04  \\
\hline
\end{tabular}
\end{table}

\section{Accretion rate}
\label{section:accretion}

Based on the previously obtained stellar parameters, HD 142527 is a
giant star with a mass typical of a late type HAeBe star and a
temperature more akin to a classical T Tauri. In order to estimate its
mass accretion rate ($\dot{M}_{acc}$), it is assumed that material in
the disc is transported to the star in the context of magnetospheric
accretion \citep[MA
  hereafter;][]{uchida1985,koenigl1991,shu1994,bouvier2007}, as this
has been shown to hold for intermediate-mass T Tauris and HAe stars
\citep{Calvet04,vink2002,muzerolle2004,mottram2007,donehew2011,Mendi11}. According
to MA, the infalling material shocks the photosphere heating the
region and releasing soft X-rays. The majority of these go on to
thermalize the surrounding, resulting in emission that is
predominately in the UV and optical. This energy is reflected by an
excess emission in the Balmer region of the spectrum
(Sect. \ref{section:balmer}) or by veiled photospheric lines in the
optical region (Sect. \ref{Section: veiling}). In addition, accretion
estimates based on the two previous observables were found to
correlate with the luminosity of several emission lines that span from
the UV to the near-IR (Sect. \ref{Section: lines}).

\subsection{Balmer Excess} 
\label{section:balmer}

\begin{figure}
\centering
 \includegraphics[width=86mm,clip=true]{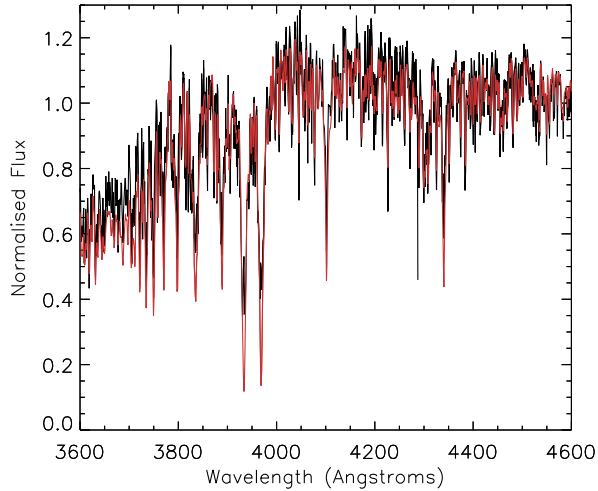}
\caption{Observed and Kurucz spectra of HD142527 are shown with in
  black and red, respectively. Both spectra were normalised to 1 at
  4000 $\rm \AA$ and the observed spectrum was adjusted to match the
  slope of the model spectrum between 4000-4600 \AA.}
\label{fig:db}
\end{figure}

The Balmer Excess ($\Delta D_B$) is given by:
\begin{equation}
\Delta D_B=(U-B)_0-(U-B)_{dered}\noindent,
\label{eqn:db}
\end{equation}

where $(U-B)_0$ is the intrinsic colour of the photosphere and $(U-B)_{dered}$ is the deredenned, observed colour. 
To perform a reddening correction the observed spectrum and Kurucz model representing the photosphere are first both normalised to 400nm. From here the slope of the observed spectrum is adjusted to match the spectra of the model between 400-460nm. This is done by measuring the difference between the observed and model and fitting the points with the \citet{cardelli1989} extinction law, which is then extrapolated to 360nm. By applying this fit to the observed spectra the slopes now match and the U band point for measurement is in the correct place. Due to the slope normalisation the B band is now the same for both object and model (see Fig. \ref{fig:db}), and the equation for the Balmer Excess simplifies to:

\begin{equation}
\Delta D_B = 2.5 {\rm log} \left( \frac{F_{int}+F_{acc}}{F_*}  \right)\noindent,
\label{eqn:donehew}
\end{equation}

where $F_{int}$ is the intrinsic flux of the star at 360nm (the model
in this case), and $F_{acc}$ is the flux due to accretion at 360nm,
such that $F_{int}+F_{acc}$ is the observed flux. This is the
expresion used in \citet{donehew2011}. In their approach $\Delta D_B$
can be measured without knowledge of a precise flux, and only the SED
shape is required.  This is appropriate for the characteristics of our
X-Shooter spectrum of HD 142527 (see
Sect. \ref{Section:data}).

The modelling used to determine $\dot{M}_{acc}$ from $\Delta D_B$
closely follows the work of \citet{calvet1998}, who applyed shock
modelling in the context of MA, and the work of \citet{muzerolle2004}
and \citet{Mendi11}, who applied the same model to HAeBe stars. We
follow the same nomenclature and methodology described in the previous
works. For a given set of stellar parameters there is only one
observable $\Delta D_B$ for each possible value of $\dot{M}_{acc}$
\citep{Mendi11}. From this, a measurement of $\Delta D_B$ can be used
to determine the accretion rate. We obtain $\Delta D_B = 0.12 \pm
0.03$. The excess emission coming from the hot spot of the
accretion column can be approximated as a blackbody function:
$F^{col}=\sigma T^{4}_{col}$, where $T_{col}$ is the temperature of
the hot spot region and $\sigma$ is the Stefan-Boltzmann constant. The
total flux from the undisturbed photosphere can also be represented as
a blackbody function: $F^{phot}=\sigma T^4_{\rm eff}$. With these
approximations the column temperature can be described as
$T_{col}=\left({\curf}/\sigma + T_{\rm eff}^4 \right)^{1/4}$, where
${\curf}$ is the inward flux of energy from the accretion column,
adopted to be $10^{12}$ ergs cm$^{-2}$ s$^{-1}$ as this has been shown
to provide appropriate filling factors ($f<0.1$) for HAeBe objects
\citep{muzerolle2004,Mendi11}. For the surface temperature of HD
142527 and ${\curf}=10^{12}$ ergs cm$^{-2}$ s$^{-1}$, the column
temperature is $T_{col} = 11800 \pm 120$ K. This corresponds to a filling
factor of $f = 0.010 \pm 0.004$, which is within the expected range, and
an accretion rate of $\dot{M}_{acc} = 1.3 (\pm 0.5) \times 10^{-7} {\rm
  M}_\odot{\rm yr}^{-1}$.  

\subsection{Photospheric line veiling}
\label{Section: veiling}

\begin{figure}
%[!hbtp]
\centering
 \includegraphics[width=86mm,clip=true]{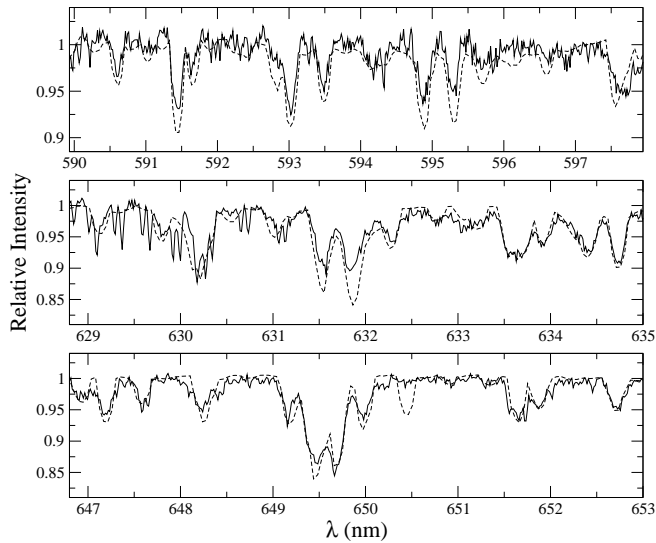}
\caption{Several regions in the optical range of HD 142527 showing
  photospheric lines. The solid line is the observed spectrum whereas
  the dashed line is the broadened Kurucz spectrum representing the
  stellar photosphere.}
\label{Figure:veiling}
\end{figure}

A second, rough estimate of the accretion rate was obtained from veiled
photospheric lines, i.e. those showing weaker absorptions than the
corresponding naked photoshere, due to the contribution of the
accretion emission. Veiling is estimated by measuring the parameter r
= (EW$^{phot}$ - EW)/EW on several photospheric absorption lines,
where EW is the observed equivalent width, and EW$^{phot}$ the
photospheric equivalent width estimated from a Kuruckz synthetic
spectrum with the same stellar parameters as HD 142527. Figure
\ref{Figure:veiling} shows excerpts of the X-Shooter spectrum in the
range 583 -- 653 nm (i.e., the range roughly covered by the Johnson R
photometric filter), where we focused our measurements. Veiling is
apparent in some photospheric lines, although it is not very intense,
with an average value r = 0.3 $\pm$ 0.1. The uncertainty comes from
the propagation of the individual EW-measurements, which are in turn
estimated from those lines with non-detected veiling, with r values
scattering about zero. The excess flux in the R photometric band is
r/(1+r) times the luminosity of the photosphere in the same band
\citep{WhiteBasri03,WhiteHill04}, which is obtained by convolving this
photometric filter with the Kurucz synthetic spectrum. The accretion
luminosity is then derived transforming the R-band excess to a total
excess by using a bolometric correction, which is adopted to be -0.4
mag. This is the value obtained by \citet{Hartigan95} for an optically
thick slab of hydrogen at 10$^4$ $K$, which is the typical temperature
of the accretion shock (Sect. \ref{section:balmer}). However, it is noted that the bolometric correction is not based on a magnetospheric accretion model but on a boundary layer one. For this reason, we assume that the accuracy of the accretion rate estimated from this method is not better than $\pm$ 1 dex. The resulting accretion luminosity, log (L$_{\rm acc}$/L$_{\odot}$) $\sim$ 0.8, can be transformed to a mass accretion rate from

\begin{equation}
\label{Eq:accretionLM}
L_{\rm acc} = \frac{GM_{*}\dot{M}_{\rm acc}}{R_*}\noindent,
\end{equation} 

deriving $\dot{M}_{acc}$ $\sim$ 3 $\times$ 10$^{-7}$ M$_{\odot}$
yr$^{-1}$.

\subsection{Line luminosities}
\label{Section: lines} 

\begin{figure*}
%[!hbtp]
\centering
 \includegraphics[width=172mm,clip=true]{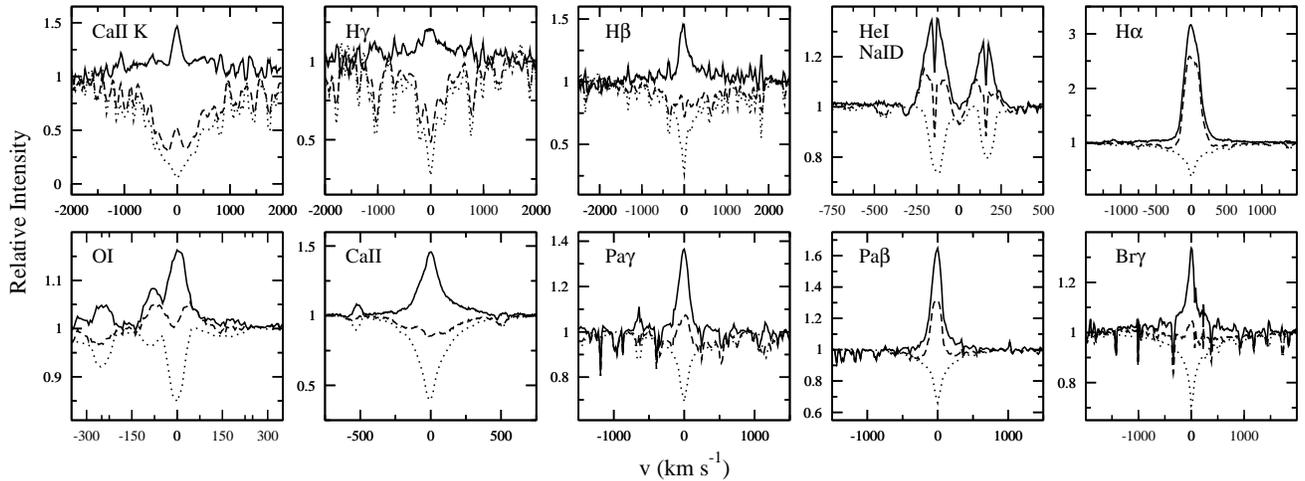}
\caption{The spectral lines used as accretion tracers are shown in the
  different panels, ordered with increasing wavelength from left to
  right and from top to bottom. Dashed, dotted and solid lines
  represent the observed, photospheric (Kurucz) and circumstellar
  (observed - photospheric) contributions, respectively.}
\label{Figure:lines}
\end{figure*}

The accretion rate was also determined from the empirical correlations
between the accretion luminosity and several emission lines that span
from the UV to the near-IR region of the spectra. These have the form
log L$_{acc}$ = a + b $\times$ log L$_{line}$, with a and b constants
that depend on the specific line considered. Accretion luminosities
are then transformed into mass accretion rates by using
Eq. \ref{Eq:accretionLM}. Figure \ref{Figure:lines} shows the emission
lines analyzed in this work, as observed in the X-Shooter spectrum of
HD 142527. Table \ref{Table:lines_accretion} lists their line
equivalent widths and luminosities, the corresponding accretion
luminosities and mass accretion rates, and the references from which
the accretion-line luminosities correlations were obtained. Line
luminosities were derived combining the circumstellar equivalent
widths, obtained by subtracting the Kurucz photospheric spectrum from
the observed one, and the continuum flux adjacent to each line,
derived from the photometry indicated in Sect. \ref{section:stellarparam}\footnote{2MASS JHK$_s$ photometry from the CDS catalogue {\tt II/246} \citep{Cutri03} was used for the lines in the NIR arm}. Line luminosities were dereddened considering
the E(B-V) value in Table \ref{Table:stelpar}, a
standard value for the total to selective extinction ratio (R$_V$ =
3.1), and the extinction law in \citet{cardelli1989}. It is noted that
the photospheric fluxes and the X-Shooter spectrum were taken on
different dates, which could introduce some uncertainty. However, it
is assumed that continuum variability is low for this object
(Sect. \ref{section:stellarparam}). Therefore, it is not expected that
results would significantly change if the photometry and the
spectrum were taken simultaneously. In addition, the empirical
correlations with the \ion{Ca}{2}K, H$\gamma$, \ion{Na}{1}D, and
\ion{O}{1}8446 lines were derived from stars with masses below the
stellar mass of HD 142527, although we assume them as equally valid
given that recent works indicate that these can be extrapolated to
more massive, hotter regimes \citep{Pogodin12,Mendi13}.

\begin{table}
\centering
\renewcommand\tabcolsep{2.5pt}
\caption{Line equivalent widths, luminosities, and accretion rates of \mbox{HD 142527}}
\label{Table:lines_accretion}
\begin{tabular}{lrrrrr}
\hline\hline
Line&EW &log L$_{line}$ &log L$_{acc}$ &log $\dot{M}_{acc}$ & Ref\\
    &(\AA{})& [L$_{\odot}$]& [L$_{\odot}$] & [M$_{\odot}$ yr$^{-1}$] &\\
\hline
\ion{Ca}{2}K (3934 \AA{})&-3.4&-1.99&0.09 $\pm$ 0.32&-7.23 $\pm$ 0.33&(1)\\
H$\gamma$ (4341 \AA{})&-1.8&-2.33&0.11 $\pm$ 0.22&-7.20 $\pm$ 0.23&(1)\\
H$\beta$ (4861 \AA{})&-3.8&-2.01&0.44 $\pm$ 0.21&-6.88 $\pm$ 0.22 &(2)\\
\ion{He}{1} (5876 \AA{})&-0.23&-3.31&0.09 $\pm$ 0.40&-7.22 $\pm$ 0.41 &(3)\\
\ion{Na}{1}D (5894 \AA{})&-0.71&-2.81&0.58 $\pm$ 0.24&-7.08 $\pm$ 0.41&(1)\\
H$\alpha$ (6563 \AA{})&-14&-1.50&0.65 $\pm$ 0.35& -6.67 $\pm$ 0.36&(4)\\
\ion{O}{1} (8446 \AA{})&-0.55&-3.05&0.89 $\pm$ 0.40&-6.42 $\pm$ 0.41&(1)\\
\ion{Ca}{2} (8542 \AA{})&-2.6&-2.38&0.40 $\pm$ 0.40&-6.91 $\pm$ 0.41&(3)\\
Pa$\gamma$ (10938 \AA{})&-3.1&-2.74&0.38 $\pm$ 0.40&-6.94 $\pm$ 0.41&(5)\\
Pa$\beta$ (12818 \AA{})&-6.5&-2.42&0.39 $\pm$ 0.40&-6.93 $\pm$ 0.41&(3)\\
Br$\gamma$ (21660 \AA{})&-7.4&-2.67&0.44 $\pm$ 0.47&-6.85 $\pm$ 0.41&(6)\\
\hline
\end{tabular}
\begin{minipage}{86mm}

  \textbf{Notes.} Typical errors for the circumstellar equivalent
  widths and dereddened line luminosities are 10$\%$ and 15$\%$,
  respectively. Error bars for the accretion luminosities consider
  those in the corresponding empirical calibrations, as indicated in
  the references provided in the last column. Error bars in the mass
  accretion rates consider those in the stellar parameters derived in this work. Data for \ion{Na}{1}D (5894
  \AA{}) refers to the sum of both D$_1$ and D$_2$ components.\\

\textbf{References.} (1): \citet{Herczeg08}, (2): \citet{Fang09}, (3):
\citet{Dahm08}, (4): \citet{Mendi11}, (5): \citet{Gatti08}, (6):
\citet{Calvet04}

\end{minipage} 
\end{table} 

According to these estimates, the mass accretion rate of HD 142527 is
in between $\sim$ 6 $\times$ 10$^{-8}$ and 4 $\times$ 10$^{-7}$, based
on the line luminosity of H$\gamma$ and \ion{O}{1}8446,
respectively. The mean accretion rate derived from the hydrogen
recombination lines has been proposed to be the most accurate
accretion rate estimate \citep{Rigliaco12}, providing 1.3 ($\pm$ 0.5)
$\times$ 10$^{-7}$ M$\odot$ yr$^{-1}$ (this uncertainty comes from the
propagation of the ones for each line). However, the averaged value
from the rest of the lines, 1.4 ($\pm$ 0.8) $\times$ 10$^{-7}$
M$_{\odot}$ yr$^{-1}$, is consistent within the
uncertainties. Therefore, it is assumed that the best accretion rate
estimate obtained through this method is the total average from all
the emission lines, providing 1.4 ($\pm$ 0.4) $\times$ 10$^{-7}$
M$_{\odot}$ yr$^{-1}$.

\section{Discussion and conclussions}
\label{section: conclussions}

\begin{figure}
%[!hbtp]
\centering
 \includegraphics[width=86mm,clip=true]{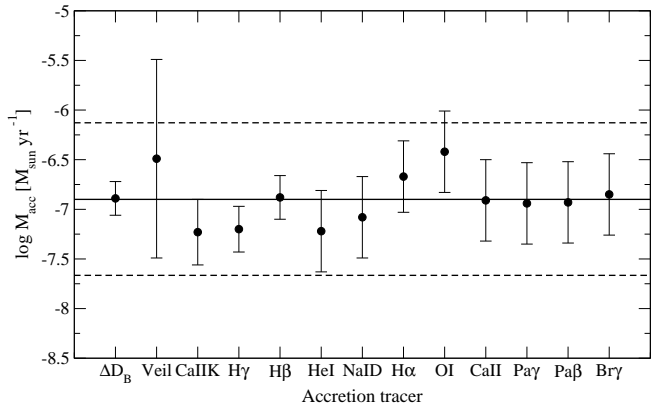}
\caption{Accretion rates obtained for HD 142527 from the indicated
  accretion tracers in the x-axis. The horizontal solid line
  represents the average accretion rate and the dashed lines the $\pm$
  3$\sigma$ deviation from the average.}
\label{Figure:accretion_all}
\end{figure}

We used a X-Shooter spectrum of HD 142527 to provide an accurate, independent determination of its
stellar parameters and accretion rate. The stellar parameters are summarized in Table \ref{Table:stelpar}, the mass accretion rates
obtained from the different tracers are plotted in
Fig. \ref{Figure:accretion_all}. All individual accretion rate
estimates are consistent to each other, given that they lie within
$\pm$ 3$\sigma$ from the mean value. Therefore, we consider this
average value as the most robust estimate of the accretion rate of \mbox{HD 142527}: 2 ($\pm$ 1) $\times$ 10$^{-7}$ M$_{\odot}$ yr$^{-1}$,
where the uncertainty is calculated from the propagation of the
individual ones. 

\citet{GarciaLopez06} estimated an accretion rate of 7 $\times$ 10$^{-8}$
M$_{\odot}$ yr$^{-1}$, assuming a distance of 200 pc and different stellar parameters than derived here. Therefore, this result
cannot be directly compared to ours. In order to make an appropriate
comparison, we first transformed the Br$\gamma$ luminosity provided in
\citet{GarciaLopez06} to the one that would be derived at a distance
of 140 pc. The same L$_{acc}$-L$_{Br\gamma}$
calibration was then applied, using the transformed line luminosity, to derive an accretion luminosity of 0.58 L$_{\odot}$,
instead of the 1.01 L$_{\odot}$ provided in that work. Finally, the
corresponding mass accretion rate from Eq. \ref{Eq:accretionLM}, using
the stellar parameters derived here, is $\sim$ 3 $\times$ 10$^{-8}$
M$_{\odot}$ yr$^{-1}$, which is a factor $\sim$ 7 lower than our
estimate. The difference cannot be associated to different procedures
but mainly to variability on a timescale of 5 yr, which is the time
elapsed between the spectrum in \citet{GarciaLopez06} and ours. The
Br$\gamma$ line in that work showed no emission whereas this can be
observed in our X-Shooter spectrum \citep[see Fig A.1 in][and our
  Fig. \ref{Figure:lines}]{GarciaLopez06}. Regarding the more recent
accretion estimate from the Pfund $\beta$ line luminosity in
\citet[][spectra taken on 2008]{Salyk13}, a mass accretion rate of 1
$\times$ 10$^{-7}$ M$_{\odot}$ yr$^{-1}$ was provided. Note that this
value, clearly above the estimate in \citet{GarciaLopez06}, refers to
exactly the same stellar parameters and distance assumed in that
work. However, that value translates again to 3 $\times$ 10$^{-8}$
M$_{\odot}$ yr$^{-1}$, once the above described correction considering
140 pc and our stellar parameters is applyed. As mentioned by
\citet{Salyk13}, their empirical calibration provides lower limit
estimates for the accretion rate, given that the accretion
measurements in \citet{Mendi11} were not considered. If these
measurements are taken into account, and using Fig. 8 from
\citet{Salyk13}, a mass accretion rate consistent with our result is
obtained. This would constrain the timescale of the observed
variability of a factor $\sim$ 7 in HD 142527 to only 2 yr.

\citet{Casassus13} obtained a gas flow rate from the outer to the
inner disk of HD 142527. According to that work, the gap-crossing
filament they found transports a mass accretion rate in-between 7
$\times$ 10$^{-9}$ and 2 $\times$ 10$^{-7}$ M$_{\odot}$ yr$^{-1}$. The
lower limit was derived from the HCO$^+$ emission, assuming that the
emitting gas is close to the critical density, n(H$_2$) $\sim$ 10$^6$
cm$^{-3}$. Because the gas is likely at a density higher than the
critical density, 10$^{-9}$ M$_{\odot}$ yr$^{-1}$ is a lower limit to
the flow through the filament. The upper limit of this flow, 2
$\times$ 10$^{-7}$ M$_{\odot}$ yr$^{-1}$, was derived from the dust
continuum emission, assuming that the fraction of continnum emission
from the filament is the same as the mass fraction of the filament
relative to the total disk mass, and that the gas-to-dust ratio is
uniform (and at the standard value) throughout the disk. However, this
assumption may not be valid if the grain abundance in the filament
differs from that of the disk at larger radii, which is the region
responsible for most of the dust continuum emission. \citet{Rice06}
have argued that pressure gradients at the outer edge of a disk gap
can act as a dust filter, letting particles smaller than a critical
size (typically larger than $\sim$ 10 $\mu$m) through to the inner
disc while holding back larger particles in the outer disc. As a
result, this process can also produce a very large gas-to-dust ratio
in the filament, which would lead to an underestimate of its gas mass,
and therefore of the accretion rate inferred by \citet{Casassus13},
potentially by a few orders of magnitude. Despite the uncertainty
introduced by this possibility, it is unlikely that the true accretion
rate through the filament would be much larger than the upper end of
the accretion rate for 2M$_{\odot}$ HAeBe stars \citep[$\sim$
  10$^{-6}$ M$_{\odot}$ yr$^{-1}$;][]{donehew2011,Mendi11}. Given the
uncertainty on the gap flow, it is difficult to state whether it is
larger, the same, or smaller than the stellar accretion rate derived
in this work, and therefore what fraction, if any, could be
intercepted by a possible planet. Nevertheless, the stellar accretion
rate we have derived assuming the same distance of 140 pc is
consistent with the upper limit in \citet{Casassus13}. Under the
hypothesis that the disk flow rate is constant -the X-Shooter spectrum
used to provide our estimates was taken 2 yr before the observations
analyzed in that work-, our stellar accretion rate would imply that
almost all gas transferred from the outer to the inner disk is trapped
by the central star, and only a small amount could be accreted onto
the planet. This would be consistent with small planetary accretion
rates, predicted to be in the range $\sim$ 10$^{-9}$ -- 10$^{-11}$
M$_{\odot}$ yr$^{-1}$ \citep[see
  e.g.][]{AyliffeBate09,Machida10}. Time-monitoring of the
simultaneous behaviour of the gas flow in the different parts of the
HD 142527 system could provide interesting results on their dynamics
and on the possible influence from a giant planet.

Finally, \citet{Biller12} \citep[see also][]{Close14} reported the detection of a low-mass stellar
companion to HD142527 separated by 88 mas from the star and at a PA of
133 degrees, i.e., not co-located with the filament reported by
\citet{Casassus12}. As reported in that work, the mass derived for the
companion is strongly dependent on the precise age of the central
object, assumed to be in the range 1--10 Myr. The tighter constraint
that we find for the age of the system indicates
that the mass of the possible companion should be in-between 0.20 and 0.35 M$_{\odot}$ \citep[see Fig. 3 in][]{Biller12}, based on
the \citet{Baraffe98} tracks.

In summary, accretion rate studies on objects with properties similar than HD 142527 are helpful not only to constrain accretion rate variability but also to give insight into possible stellar/planetary companions.

\acknowledgments

Based on observations made with ESO Telescopes at the La Silla Paranal
Observatory under programme ID 084.C-0952. B. Montesinos is supported 
by Spanish grant AYA2011-26202. S.D. Brittain acknowledges support for this work from the National Science Foundation under grant number  AST-0954811. The authors are grateful to Jes\'us 
Ma\'{\i}z-Apell\'aniz for fruitful discussions concerning the photometry
of this star. The authors thank the anonymous referee for his/her useful
comments on the original manuscript, which helped us to improve the paper.\\


%% in parentheses, after the keyword, but they will not be verified.

{\it Facilities:} \facility{VLT:Kueyen}

%Appendix, if necessary
%\clearpage
%\appendix

%\section{Appendix material}

\end{document}